# *Current state of the Sonix – the IBR-2 instrument control software and plans for future developments*

A.S. Kirilov

FLNP of JINR

141980, Dubna, MoscowReg., Russia

kirilov@nf.jinr.ru

The Sonix is the main control software for the IBR-2 instruments. This is a modular configurable and flexible system created using the Varman (real time database) and the X11/OS9 graphical package in the OS-9 environment. In the last few years we were mostly focused on making this system more reliable and user friendly.

Because the VME hardware and software upgrade is rather expensive we would like to replace existing VME + OS9 control computers with the PC+Windows XP ones in the future. This could be done with the help of VME-PCI adapters.

## *Contents*

- Instruments with SONIX.
- To achieve more stable work.
- Making the Sonix more user friendly.
- On-line visualization.
- Using VME-PCI adapters.
- References.

## *Instruments with SONIX*

The SONIX system/1/ is modular, highly versatile, configurable software complex for neutron instruments on the basis of the X11/OS9 graphical package. It was designed at FLNP originally for the NSHR spectrometer and then ported to the most of IBR-2 instruments.

In the following table the current situation of use SONIX at IBR-2 beamlines/2/ is summarized.

| Beam | Instrument | Instrument description | Commission | Commentary |
|---|---|---|---|---|
| 4 | YuMO | SANS spectrometer | 1999 | |
| 6a | DN-2 | multipurpose diffractometer | 2000 | |
| 7a | SKAT | texture diffractometer | 1997 | |
|    | EPSILON | stress diffractometer | 1998 | |
|    | NSHR | texture diffractometer | 1995 | |
| 7b | NERA-PR | Multi-crystal inverted geometry spectrometer | 1998 | |
| 8 | SPN | spectrometer of polarized neutron scattering | 2002 | testing |
| 11a | FSD | Fourier stress diffractometer | 2002 | testing |
| 12 | DN-12 | high-pressure diffractometer | 2001 | |
|    | SAX | X-ray diffractometer | 1999 | |
|    | DRON | X-ray diffractometer | 2001 | |

## *To achieve more stable work*

Making spectrometer software work more stable was one of our main long term goals. The Os-9 system itself is very stable. Unfortunately, Internet Support Package for this system is not so stable. There are two kinds of errors - nonfatal and fatal. The nonfatal errors block i/o activity of the computer via Internet but do not affect the rest of the operating system and tasks which unrelated to the network. The fatal errors hang the system completely. The occurrence of errors depends on the concrete place of the VME computer in the network. It also depends on the traffic intensity. At some instruments where were no errors at all. At the others in the worst case where were one nonfatal error per day and one fatal error per week. To make effects of these ISP errors negligible we preserve some special structural features.

- We organize the complex so that the modules which are really responsible for the measurement (we call them - residents) are separated from the so-called interfaces which are used for the user control (local and remote). Both residents and interfaces communicate through the variables of the real time database created and maintained by the Varman/3/. The residents do not deal with the network so the core of the measurement software is insensitive to nonfatal ISP bugs. Of course, in case of these bugs the user loses remote control over the system but he can restart the computer if he really needs to renew the control.

- At every moment the contents of the Varman database completely represents the current state of the measurement process at the instrument. The database can be saved and restored. This feature enables an automatic restart of the measurement if the system is hung up after a fatal ISP error. This is implemented using a watchdog device incorporated at the E17 VME processor board/4/. The operating system will be restarted automatically in 1.6 sec. Then all residents and the database will be reloaded. Thereafter the measurement will be continued from the nearest point specified by user in the script file.

## *Making the Sonix more user friendly*

Another main long term goal was to make user interface more friendly. Our script language was one of the main points of criticism. The interpreter is a very flexible tool in creating various measurement procedures. But our experience shows that occasional users (visitors) may have problems when creating script files. To say nothing of typing errors they sometimes use "wrong" editor, "incorrect" ftp programs to transfer this file to VME, etc. Thus, several steps were made to enhance script language:

- The syntax of procedure calls was changed. Now they look similar to macro-commands.
- The question statement was added to edit parameter value interactively. This feature decreases the necessity of script editing.
- The design of macro-command library suitable for requirements of a specific instrument was also changed a great deal to simplify script creation.

The script is more suitable for the experienced people like instrument responsibles. Visitors certainly need a special kind of interface programs with a rather strict dialogue protecting them from typing and logical errors. The attempt to create such a program was made for one of the most popular, charged and sophisticated instrument at FLNP – the YuMO SANS spectrometer. This program (Sans Editor) runs on PC. It assists the user to create a script and loads it into the YuMO VME control computer.

```
;*************************************************************
;Measurements:15.05.2002,P. Balgavy
;Valentin'schamber
;(Hexane+Dodecane) 500ulperpendicullar to beam
```

```
; ********** 1mm cuvettes ********************************
;++++++++++++++++++++++++++++++++++++++++++++++++++++++++++
;
auto_test
;
Motor:open_prot
Tofa:open_prot txt/pb160502a.txt
Temp:open_prot txt/pb160502at.txt
Unipa:open_prot txt/pb160502au.txt
Motor:getpos
;
usf_set(Balgavi,Hexane,PB160502a)
;
;Task for checking of unipa-parameters
;
uni_start(test)
;++++++++++++++++++++++
#set @filename PB160502a
Tofa: file @filename
;+++++++++++++++++++++++++++++++++++++++++++
shut_set(vanady1_2det,outbeam)
shut_set(vanady1_1det,outbeam)
shut_set(vanady2_2det,outbeam)
shut_set(vanady2_1det,outbeam)
;
;++++++++++++++++++++++
temp_ist(1.0,2.0,test,25)
Tofa:flagoff temperature
;---------------------------------------------------------
;start with measuring of sample number 11
meas_2sh(vanady1_1det,vanady1_2det,2000,1000,1,11,#.$09)
;---------------------------------------------------------
Unipa: stop
; ----- eof -----
```

Fig.1   The YuMO sample script program

Figure 1 shows the YuMO sample script and Figure 2 presents the Sans Editor program interface.

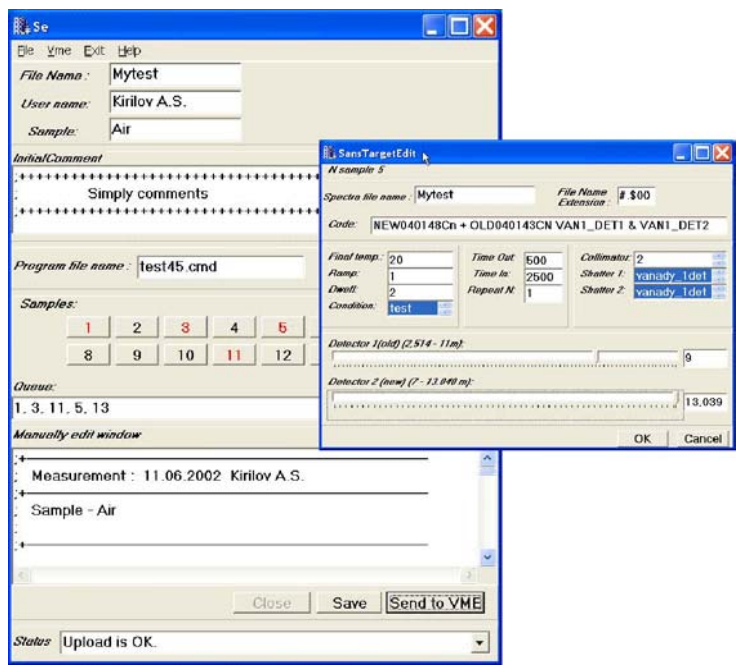

Fig.2   The YuMO script editor

Of course, the interface of this editor is a result of the compromise. It can create scripts only for a typical experiment. This program is instrument dependent. At the moment it cannot be configured by users.

This is our first experiment in this field. When we gain more experience on this subject we shell be able to propose more universal software. We intend to create similar programs for the other IBR-2 instruments also.

### *On-line visualization*

The Sonix includes the VSP program for on-line visualization of one-dimensional spectra. Like other Sonix components VSP is completely implemented on the VME-side. Restrictions on E17 capacity and network speed force us to develop special on-line visualization program for instruments with PSD to minimize traffic. On our first intention special VME daemon working in tight cooperation with DAQ program of the Sonix on the external request must sample and compress data in the intermediate file on the VME side. Then another daemon on the PC transmits this file and creates a picture.

This set of modules named MAKS/5/ was developed for the DN-2 instrument. The experience with MAKS has shown us that this program is useful not only for PSD instruments but also for other setups because it has less traffic than VSP and VME computer work becomes more stable. Thus this program was installed at some other instruments of IBR-2.

For the last year practical capacity of FLNP local network has increased greatly. We have investigated MAKS operation timing again and figured out that with high transmission speed the compressing of data on the VME side has no sense. At the moment, a new version of MAKS with "direct" data read out from the histogram memory is under development.

### *Using VME-PCI adapters*

VME system upgrade both in hardware and software is too expensive. The cheapest computers suitable for instrument control are PCs. In our opinion the natural way of combining PC control with the existing VME custom-designed hardware is to use VME-PCI adapters/6/. This idea has come to FLNP from ILL. At the moment the first system of this kind is being created at the NERA-PR multi-crystal inverted geometry spectrometer. For this purpose PC (P4 – 1.3GGz, 256Mb) was connected to the existing VME system based on E17 with the help of Model 617 adapter form SBS. On the PC the Windows XP was installed.

At the first stage we decided to replace only the interface part of the Sonix, preserving the existing resident part. Both parts: new interfaces at PC side and old residents at VME side communicate through the 128K dual port memory of the adaptor.

The Varman has already been ported to the Windows. It is used as a communication environment for interfaces.

We hope that this software will be ready for testing at the autumn IBR-2 measuring period. Our main goal will be to evaluate the reliability of the complete PC software system (our software + Windows XP).

### *Acknowledgement*

The author is grateful to all his colleagues from the Spectrometers Support Department of FLNP.

### *References*

1. http://nfdfn.jinr.ru/~kirilov/Sonix/sonix_index.htm

2. http://nfdfn.jinr.ru/fks/pssol.html

3. Sckipper M.N. The Real-Time Database Solution at IRI. Proc. of the DANEF'97 (June 2-4, 1997, Dubna, Russia), E10-97-272, JINR, Dubna, pp. 288-294.

4. EUROCOM-17-5xx.Dual 68040 CPU Board with Graphics. Hardware Manual. Revision 1A. V-E17.-A995. ELTECElectronickGmbH.

5. http://nfdfn.jinr.ru/ibr2_research_workshop2002.htm

6. http://www.sbs.com/computer/products/cp_pci_vme_hp.shtml